# Cloud Computing: a Prologue

Sultan Ullah, Zheng Xuefeng

School of Computer and Communication Engineering, University of Science and Technology, Beijing China.

ABSTRACT— *An emerging internet based super computing model is represented by cloud computing. Cloud computing is the convergence and evolution of several concepts from virtualization, distributed storage, grid, and automation management to enable a more flexible approach for deploying and scaling applications. However, cloud computing moves the application software and databases to the large data centers, where the management of the data and services may not be fully trustworthy. The concept of cloud computing on the basis of the various definitions available in the industry and the characteristics of cloud computing are being analyzed in this paper. The paper also describes the main cloud service providers and their products followed by primary cloud computing operating systems.*

## I. INTRODUCTION

Cloud Computing represents the future of information age. It is not an instant, turned out to be new technology; rather it is the evolution of distributed computing, parallel computing and grid computing. Due the results of this hybrid evolution, the applications can be extended via the internet. There are many well known information technology companies which are the pioneers in offering cloud computing infrastructure or cloud computing platform, such as Sun's Cloud infrastructure, IBM's "Blue Cloud", Google's cloud computing infrastructure, Microsoft's Azure cloud platform, and Amazon's Elastic cloud.

In fact cloud computing is a hot research area in the academia, and under the joint venture of the major information technology companies and academia is consistently working on the improvement of user friendliness, security and many other aspects of cloud computing. It is worth mentioning that with the improvement of network software and network speed, the development of cloud computing is very fast. One can easily predicts that in the next 15 to 20 years cloud computing will become a key factor of business for the entire information technology industry. Cloud computing is not yet mature, the security of cloud computing will determine the success, development and growth of this new technology.

The left over paper contains the different concept of the cloud in section II. The service model of the cloud is discussed in section III. In section IV the characteristics of the cloud are presented. In section V different cloud service provider and their products are discussed. Section VI contains the operating system for the cloud, section VII presents the conclusion of the paper followed by references.

## II. THE CONCEPT OF THE CLOUD

Since the emergence the term "Cloud Computing", major IT companies and academia give different definitions of the cloud computing. Everyone has define cloud computing from different angles, but the real essence of cloud computing is that it use internet to provide services and applications for business and scientific use.

Cloud Computing can be defined in the following ways:

The concept of cloud computing was first presented at the end of 2007 by IBM Corporation in its Technical White Paper [1], and pointed out that the cloud is a virtualization of computer resources. On one hand the paper describe the services provide by cloud computing, for examples the system platform, application infrastructure, and on the other hand it describe the applications access can be extended via internet. The user only needs a standard web browser and internet access devices to access the cloud computing application over the internet [1].

Michael Armbrust [2] in his paper above the cloud: a Berkley view of the cloud computing states that cloud computing refers to both the applications delivered as services over the internet and the hardware and systems software in the datacenters that provide those services themselves have long been referred to as Software as a Service (SaaS). The datacenter hardware and software is what we call a cloud. When cloud is made available in a pay – as – you – go manner to general public, we call it a Public Cloud; the services being sold is Utility Computing. We use the term Private Cloud to refer to internal datacenters of a business or other organizations, not made available to everyone. Thus, cloud computing is the sum of SaaS and Utility Computing, but does not include Private Clouds. People can be users or providers of SaaS or Users or providers of Utility Computing.

Wikipedia [3] defines cloud computing is to provide computing as service instead of computing as product, where shared resources i.e software, information, and other devices are provided to computer user as a utility over a network. These resources are dynamic, centralized and easy to expand. The cloud generally refers to large scale server clusters.

Cloud Computing as defined by the US National Institute of Standards and Technology (NIST) states that:"Cloud computing is a model for enabling convenient, on-demand network access to a shared pool of configurable computing resources (e.g., networks, servers, storage, applications, and services) that can be rapidly provisioned and released with minimal management effort or service provider interaction. This cloud model promotes availability and is composed of five essential characteristics, three delivery models, and four deployment models." [4].

### III. THE SERVICE DELIVERY MODEL OF THE CLOUD

In cloud computing the primary task is to achieve XaaS. From an architectural point of view, the underlying layer of cloud consist of cloud hardware and form the basis of IaaS (Infrastructure as a Service), then Platform as a Service (PaaS) and Software as a Service (SaaS). These three layers not only contain the information required by the user, but also define a new application development model. Since cloud computing is an infrastructure, it is divided into three layers [5][6] which are as under:

- Infrastructure as a Service (IaaS)
- Platform as a Service (PaaS)
- Platform as a Service (PaaS):

#### A. Infrastructure as a Service (IaaS)

First layer is the IaaS. It includes the basic infrastructure for an application to be run e.g. all the hardware (processor, storage and servers etc) and all the software (network, processor speed etc). These resources are provided as a service. Cloud service provider demands required infrastructure (hardware and software) and in return a virtual machine is provided to him having all the capabilities required by him.

#### B. Platform as a Service (PaaS)

Second layer is the PaaS. It includes all the facilities to develop an application and its execution on appropriate infrastructure.

#### C. Software as a Service (SaaS)

Third and last layer is the SaaS. It is the full fledged software (application) that is delivered to the end developed using the infrastructure and platform provided above three layers.

### IV. THE CHARACTERISTICS OF A CLOUD

According to the above mentioned different concepts of cloud computing, therefore it is concluded that the cloud has the following characteristics;

A. *Internet – Centric:* The most important characteristic of the Cloud computing is that the services which is offered by the cloud can be accessed through internet.
B. *Virtualization:* In cloud environment, networks, servers, storage, constitute a pool of virtual resources.
C. *Reduce the Terminal Equipment Requirement:* The users of the cloud only need web browsing software and internet access devices to access the application of the cloud, and thus reduces the requirement for the terminal equipment.
D. *Ease of use for the User:* The user does not need any concern about the details that what is inside a cloud environment, and does not need to know the relevant expertise how to manage or control the underlying infrastructure. He only needs to know the network access to the service.
E. *Scalability:* Cloud Computing can expand seamlessly, as the dynamic expansion is possible from a small scale to large scale cluster.
F. *Facilitate Data Sharing:* Cloud computing environment is a distributed environment, so different user simply connect to the internet can share the same data.
G. *Pay per Use:* The most important characteristic of cloud computing paradigm is pay per use policy. If the user is using the services then he is to pay, if not then there is no need to pay.

### V. MAJOR CLOUD SERVICES PROVIDERS AND THEIR PRODUCTS

The list presented by Lynch after the analysis, shows that Google, IBM, Microsoft and Amazon topped the list in terms of service delivery of cloud computing [7]. Google is the largest user / provider of cloud computing services [8]. Google introduced Google App Engine which allows the developer to write Python – Based applications. In addition, Google also provide Storage server up to 500 MB of Google infrastructure for storage, bandwidth and CPU resource usage free of charge.

The Google File System is the most important product of Google, which is a distributed file system and support MapReduce programming model and the BigTable support which is the large scale distributed database system[9][10][11]. The current target of the Google is to attract the Information Technology department of the company instead of the individual users to use its cloud services. Google and IBM has a joint venture in the united states on the name of Google – IBM cloud. Google and IBM came together the Google Machines, the IBM system X servers, at the BladCenter running Linux and Xen as virtualization software and the Apache open source Hadoop framework [11].

In 2007, IBM launched "Blue Cloud" program. The "Blue Cloud" is a set of hardware and software platform that will be used on the internet, which is an extension to the enterprise platform. Blue Cloud was the extensive use of the IBM first large scale computing, which combine the use of IBM's own hardware and software system, as well as the service technology to support the open standard and open source software. The cloud infrastructure of the Blue Cloud is based on IBM Xen, using Linux as operating system, PowerVM as virtualization software and for the system as well as for the distributed computing accept open source framework for the Hadoof [11]. The advantage of the IBM "Blue Cloud" product is they introduced the system of server on x86 chips, while in the next step they will be based on IBM system Z mainframe cloud.

In October 2008, Microsoft launched cloud computing platform known as Windows Azure. The underlying technology of the Azure Services Platform is Microsoft's new cloud operating system Windows Azure [13]. This is the foundation of Microsoft's next generation network services. The Microsoft's strategy is to cover individuals, corporate and third party developer and to bring them to Microsoft's cloud services.

Amazon launched its cloud services under the name Elastic Compute Cloud [14]. It provides Amazon Web Service (AWS), Simple Storage Service (S3), simple queuing service (SQS) and Database Service (Simple DB) [15]. The user needs to pay to Amazon in order to avail these services. Elastic Compute Cloud (EC2) provides access to different types of Xen style virtual server. The Amazon also offers two permanent data storage service i.e S3 and Simple DB. The Amazon charges fee for its storage services, bandwidth and CPU resources and so on. They storage and bandwidth capacity charges vary and CPU resources are charged on long run time.

## VI. MAJOR CLOUD OPERATING SYSTEMS

The Operating System is the most important system software not only for the personal computers but it is of much importance for the cloud computing. The cloud operating system is not only system software which performs the management of the cloud but it also provides a platform for a variety of hardware and software resources and to provide the user with the cloud interface. There are two main kinds of cloud computing operating system.

### A. Lightweight Browser Based Operating System

This type of operating system works mainly through the browser. Browser provides the users with a way to access the cloud computing platform. Browser performs a similar function to the traditional operating system in order to complete a task. Therefore this type of operating system is a light weight, and manifestation of Software as Service (Saa) of cloud operating system. The typical examples of this type of system include eyeOS [16], icloudOS [17] and Microsoft's Live Mesh [18]. These relatively simple browsers based lightweight operating system, and the way they access to cloud computing platforms, they cannot be regarded as true cloud operating system.

### B. Heavyweight Hardware Based Operating System

The functionality of these operating system is to drive hardware, manage cloud computing platform for a variety of hardware and software resources, and to provide the user with a transparent view of services of the cloud computing. This group of operating system consists of Google App Engine, Microsoft Windows Azure, VMware VDC-OS, Amazon's EC2, and MIT has recently proposed cloud operating system FoS.

1) *Google App Engine [12]:* Google App Engine lets you run web applications on Google's infrastructure. It reflects the idea of platform as a service (PaaS), and spans the cloud computing platform for multiple servers and datacenters to virtual applications. App Engine applications are easy to build, easy to maintain, and easy to scale as your traffic and data storage needs grow. With App Engine, there are no servers to maintain: You just upload your application, and it's ready to serve your users.

2) *Microsoft's Windows Azure [13]:* This operating system us developed by Microsoft to supports its cloud computing platform operating system. Azure is basically a platform for developers of web applications. The operating system provide cloud platform to provide online services, basic storage and management of the cloud platform. Azure allows the user to Live Services, NET services, SQL Services, Share Point Services and Dynamic CRM. Azure allows the user to developed and implement online services on the cloud platform. At the present most of the services are provided by Microsoft online to provide with the support of windows Azure cloud computing service, such as SQL Azure. Google App Engine as compared with Windows Azure, the adaptive range of Azure is much greater, but due to Azure born with Microsoft trade mark, specific application can be constructed.

3) *VMware VDC-OS [19]:* The Virtual Datacenter OS allows businesses to efficiently pool all types of hardware resources - servers, storage and network – into an aggregated on-premise cloud – and, when needed, safely federate workloads to external clouds for additional compute capacity. Datacenters running on the Virtual Datacenter OS are highly elastic, self-managing and self-healing. With the Virtual Datacenter OS from VMware, businesses large and small can benefit from the flexibility and the efficiency of the "lights-out" datacenter. IBM "Blue Cloud" computing strategy related to the allocation of resources through virtualization. The virtualization technology reflects the Infrastructure as a Service (IaaS) concept. From the original intentional view of the cloud computing, VDC-OS is the most successful and representative of the cloud operating system. But the VDC-OS does not provide the communication mechanism between different virtual machines, and between user applications on different virtual machines. So, interoperability is difficult in this OS.

4) *Amazon Elastic Compute Cloud [13]:* Amazon Elastic Compute Cloud (Amazon EC2) is a web based operating system which provides service that provides resizable compute capacity in the cloud. It is designed to make web-scale computing easier for developers. It has similar idea of IaaS when compared with the VM-VDC. The communication between different machines is more complex.

5) *MIT'S FoS [20]:* It is developed at the MIT Carbon Research Group in 2010. FoS is a new operating system targeting multicore, manycore, and cloud computing systems with scalability as the primary design constraint, where space sharing replaces time sharing to increase scalability. This operating system is different from the other cloud computing operating systems like VM-DC, EC2. It presents a single view to all users, and provide user with a unified programming model for the expansion of operating system services.

## VII. CONCLUSION

It can be deduced in accordance with the above mentioned cloud computing characteristic and different definitions of the cloud computing concept; it can be seen as fusion of virtualization and web 2.0 technologies. For instance it present a super computing model, the cloud server is connected over a network connection, a virtual machine approach which constitute a virtual resource pool, with a superior computing power having the ability and capacity of data storage and sharing via the internet to the user.